\documentclass[showpacs,showkeys]{revtex4}
\usepackage[dvips]{graphicx}
\usepackage{epsfig}
\begin{document}

\title{Harmonic oscillators in the Nos\'e - Hoover thermostat}

\author{V.L. Golo$^1$}
\email{golo@mech.math.msu.su }

\author{Vl.N. Salnikov$^1$}
\email{vladsal@direct.ru}

\author{K.V. Shaitan$^2$}
\email{shaitan@moldyn.ru }

\affiliation{$^1$ Department of Mechanics and Mathematics \\
     $^2$ Department of Biology \\
     Moscow University \\
     Moscow 119 899, Russia \\  }

\date{January 29, 2004}

\begin{abstract}
   We study the dynamics of an ensemble of non-interacting harmonic oscillators in a nonlinear
   dissipative environment described by the Nos\'e - Hoover model.
   Using numerical simulation we find  the histogram for total energy, which agrees
   with the analysis of the Nos\'e - Hoover equations effected with the method of
   averaging. The histogram  does not  correspond to Gibbs' canonical distribution.
   We have found oscillations at frequency proportional to $\sqrt{\alpha/m}$,
   $\alpha$ the dissipative parameter of thermostat and $m$ the characteristic
   mass of particle, about the stationary state corresponding to equilibrium.
   The oscillations could have an important bearing upon the analysis of
   simulating molecular dynamics in the Nos\'e - Hoover thermostat.
\end{abstract}


\maketitle

\section{\label{sec:intro}Introduction}

The Nos\'e - Hoover model, \cite{nos},\cite{hoover_holian}, is
widely used in molecular dynamics for simulating a system's
behavior at constant temperature.  The central idea of the model,
that is the introduction ancila dynamic variables to control
kinetic energy, admits of various implementations. In the most
simple and exploited form it amounts to considering a minimal
non-linear extension of the original equations  for the system. It
should be noted that hamiltonian versions of the model drew
considerable attention, \cite{hoover}.

In the present paper we study the Nos\'e - Hoover model in the
form generally employed in molecular dynamics, that is a system
constructed from initial hamiltonian equations, i.e. Newton's
second law, by employing non-linear dissipative terms on their
right-hand sides and an additional equation for the dissipation
constant,$\gamma$, which is allowed to vary in time, so that the
equations of evolution read
\begin{eqnarray}
  m_i\ddot{\vec r}_i &=& -\frac{\displaystyle \partial}{\displaystyle
  \partial \vec r_i} U(\vec r_1,\vec r_2,\ldots,\vec r_N)
  - \gamma \dot{\vec r}_i
    \label{N-H1}    \\
  \dot \gamma &=& \alpha \left(
  \frac{\displaystyle 2}{\displaystyle 3k_bTN}
  \sum_{i=1}^{N} \frac{\displaystyle m_i \dot{\vec r}_i^2}{\displaystyle 2}
  -1 \right) \nonumber
\end{eqnarray}
In this setting the Nos\'e - Hoover model is a hamiltonian system
with dissipation. For small values of $\alpha$ the dissipative
effects can be treated within the framework of perturbation
theory.

Considerable criticism has been levelled at the Nos\'e - Hoover
approach, (see paper \cite{Als} and references therein), because
it runs across difficulties in providing the correct thermodynamic
behavior for simple harmonic systems,  but it is still generally
accepted that its efficiency improves with the increase of
complexity and dimension of the simulated system,\cite{Als}. In
fact, from the point of view of dynamical theory it models a
non-linear time-dependent dissipative environment and, therefore,
has some proper physical interest.

In this paper we focus our attention on ensembles of harmonic
oscillators; the importance of such systems follows from the fact
that among these, is the harmonic lattice, familiar  in the
theories of solids and molecules. The potential energy of a
harmonic lattice is given by the quadratic form
\begin{equation}
  U(\vec r_1,\vec r_2,\ldots,\vec r_N) =
    \sum_{i,j=1}^{N} \sum_{l,k=1}^{3} \lambda_{ij}^{lk} r_{i}^{l} r_{j}^{k}
    \label{pot1}
\end{equation}
in which $r_i^l$ is the $l$-th coordinate of the $i$-th particle.
To see the symmetry properties of the model we may  cast
Eq.(\ref{N-H1}) in the matrix form
$$
   m_i \ddot{r}_i + (\Lambda r)_i + \gamma \dot{r}_i = 0
$$
in which $\Lambda$ is the matrix of force constants,
$\lambda_{ij}^{lk}$. It can be transformed by an appropriate
orthogonal transformation, $R$, to the diagonal form
$$
  U(w_1,w_2,\ldots,w_{3N}) =
	\sum_{i=1}^{3N} \lambda_{i} w_{i}^{2}
$$
Assume that all $m_i$ are equal, $m_i = m$, and  let  $R$ be the
matrix of the orthogonal transformation  mentioned above, and
$w$ an $N$-dimensional vector of coordinates with respect to
the new coordinate system determined by $R$. Since $R(t) = R$ is
constant, and $\gamma$ is an invariant of orthogonal
transformation, we may cast Eq.(\ref{N-H1}) in the form
$$
  R\ddot{w} + \Lambda Rw + \gamma R \dot{w} = 0,
$$
so that  the equation acquires the form
$$
\ddot{w} +(R^{T} \Lambda R)w + \gamma \dot{w} = 0
$$
in which the matrix $R^{T} \Lambda R$ is the diagonal one. Thus,
we have transformed the original problem of harmonic lattice to
that for a set of harmonic oscillators, which do not interact with
each other. The latter problem is more tractable, from analytical
point of view.

At this point it is worthwhile to notice that problems of
molecular dynamics involves dynamical systems of extremely high
dimension, and this circumstance brings about specific difficulties
for numerical simulation. The best approach to the problem is to use analytical
methods in conjunction with the numerical ones, especially in
cases like the Nos\'e - Hoover model, where systematic
investigation of high dimensional problems is particularly
interesting, (see \cite{Als}). For this end we use the method of
'windows' worked out for the needs of the relaxation dynamics of
spin in superfluid $^3He$, (see the review article \cite{golo}) to
obtain a general picture of the Nos\'e - Hoover dynamics for an
ensemble of harmonic oscillators. We show that it is characterized
by the presence of oscillations round the stationary solution
corresponding to an equilibrium, for  which the phase-space
sampling is different from the normal law.

\section{\label{sec:histogram}  Stochastic properties of the system}

The numerical analysis of the Nos\'e - Hoover model given by
Eqs.(\ref{N-H1}), which is a high-dimensional non-linear
dissipative system, is a serious challenge, and, in fact, it is
generally confined to low dimensional situations,(see paper
\cite{posch}, in which the case of two oscillators is considered).
In treating high-dimensional problems the key point is the wise
choice of output variables. In the present case it is dictated by
the physics of the problem, and taking into account the structure
of the Nos\'e - Hoover model, that is its being a hamiltonian
system with  dissipation, we  employ, to the effect, the total
energy of the system $E$,
$$
    E = E_{kin} + U,
$$
and the dissipative variable $\gamma$. Directing the output in
plane $(E - \gamma)$, we obtain a kind of two-dimensional window
on the phase-space of the model, which has dimension $2 N + 1 $,
$N$ is the number of oscillators.  Since we aim at studying the
situations in which $N$ is large, the  two-dimensional
reduction is of primary importance. Next, one should look for the
distribution law for the the system in phase-space, and one could
expect that it should be  either the microcanonical or the Gibbs
one; the first is characterized by its being centered on a
particular value of energy, $E_0$, whereas the latter by its being
of the characteristic bell-shape. Nothing of the sort happens.

\begin{center}
 \begin{figure}[h]
	 \includegraphics*[width=3in, height=2in]{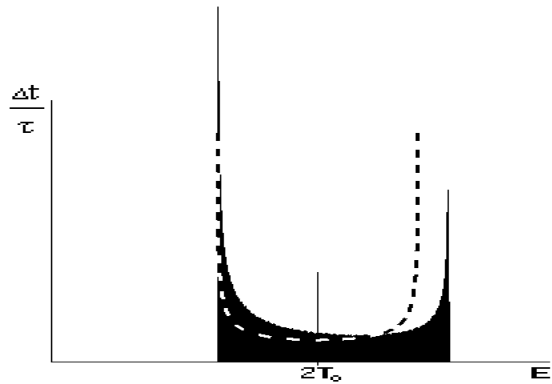}
     \caption{\label{hyst1p}
       Energy regions of the phase-space against visiting time;
	   the number of oscillators $N=1000   , \alpha = 0.01$.
	   $T_0 = Nk_bT/2$. Dashed line corresponds to
	   Eq.(\ref{arcsin}) provided by the averaged equations.
	 }
 \end{figure}
\end{center}

\begin{center}
 \begin{figure}[h]
	 \includegraphics*[width=3in, height=2in]{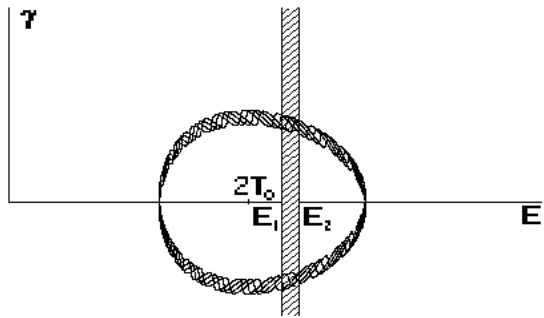}
	  \caption{\label{build}
	Trajectories visiting regions ${\cal R}_k$ in ($E$ - $\gamma$)
	window. $N = 1000$, $\alpha = 0.01$,
	$T_0 = Nk_bT/2$.
			  }
 \end{figure}
\end{center}

To find the distribution law we consider the partition of the
phase-space in regions
\begin{equation}
  {\cal R}_1, {\cal R}_2, ... {\cal R}_k, ...{\cal R}_L \label{regions}
\end{equation}
corresponding to the energy intervals $ E_k \le E \le E_{k+1} $
assumed to be of equal size, and compile a record of periods of
time $   t_1, t_2, ... t_k, ... t_L $ which the system spends in
regions (\ref{regions}); the total time of simulation reads
$$
  t_{total} = \sum_{i=1}^{L} \, t_i
$$
The frequencies for the system's visiting the regions are given by
the equation
\begin{equation}
   p_k = \frac{t_k}{\displaystyle{t_{total}}}, \quad k = 1,2, \ldots L  \label{visiting}
\end{equation}
It is convenient to use a representation for the set of
frequencies by means of histogram, that is rectangles whose widths
represent the energy intervals (\ref{regions}) and whose heights
represent corresponding frequencies. It is worth noting that the
partition of the phase-space in the energy regions (\ref{regions})
can be effected in a more graphic form with the help of the window
$(E - \gamma)$ on the phase space.  In fact, the numerical
simulation gives the picture of the system's motion as a thin
ring, the times $t_k$ are those spent in the bands $E_k \le E \le E_{k+1}$.
Therefore, the frequencies (\ref{visiting}) could be found with
the two-dimensional window, $E - \gamma $, see FIG.(\ref{build}).

Recall that we shall consider a set of N {\em one-dimensional}
harmonic oscillators.

\begin{center}
 \begin{figure}[h]
	 \includegraphics*[width=3in, height=2in]{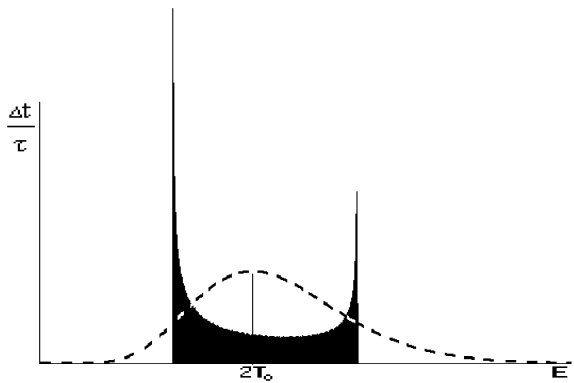}
      \caption{\label{maxwell1p}
      Bars, energy distribution, obtained from numerical experiment.
	  Curve, Gibbs' distribution for the set of harmonic oscillators.
	  $N=1000   , \alpha = 0.01$,
	  $T_0 = Nk_bT/2$.
     }
 \end{figure}
\end{center}

By following the prescription given above we obtain the histogram
given in FIG.(\ref{hyst1p}), describing the probability
distribution $\rho$ for the system in the phase-space . It is
quite different from the microcanonical one or the Gibbs one. It
is important that the window $(E - \gamma)$  provides a means for
elucidating the form of the fluctuations round the stationary
state given by the equations
$$
   \gamma =0, \quad E = 2 T_0, \quad E_{kin} = T_0
$$
The circular motion seen from the window $(E-\gamma)$ is
characterized by a mean angular velocity given by the law
$\sqrt{\alpha / m}$ to within one-thousandth. The fact agrees with
the histogram given in FIG.(\ref{hyst1p})(see
Section \ref{sec:averaging} for the details).
For the case of ideal gas the oscillation law $\sqrt{\alpha/
m} $ was found in \cite{posch} (see Eqs.(5) and (8) of paper
\cite{posch}). It should be noted that the shape of the histogram
depends on the amplitude of the oscillations, that is the size of
deviations in the energy $E$ from the stationary value $2 T_0$
determined by the temperature parameter of the model. In the next
section we shall find the distribution using averaging method, and it
is worth noting that the numerical results are in good agreement
with those given by the averaging, as is seen from
FIG.(\ref{maxwell1p}). The
distribution in energy that should correspond to Gibbs' canonical
ensemble for the set of harmonic oscillators at temperature $T$,
is given by the equation
\begin{equation}
	   d \rho = c_N \, \frac{E^{N-1}}{\displaystyle{(2 \pi k_b T})^N} \,
	   e^{\displaystyle{- \frac{E}{k_bT}}} \, dE
       \nonumber
\end{equation}
and is totally different from FIG.(\ref{hyst1p}). The discrepancy
between the obtained and expected distributions indicate that the
Nos\'e - Hoover model describes a kind of non-linear dissipative
system having special properties.

\section{\label{sec:averaging} Averaged system}

The simulation of the last section, which was thoroughly checked
by calculating with different algorithms and comparing their
results, might nonetheless be subject to artifacts and errors. In
this respect, it is important that analytical means capable of
verifying section \ref{sec:histogram}, have been used and resulted
in agreement with the numerical work.

Let us notice that according to the prescription described in
Introduction (see Eqs. (\ref{N-H1})) the ensemble of $N$ harmonic
oscillators confined to the Nos\'e - Hoover thermostat is
described by the system of equations
\begin{eqnarray}
  \ddot{x}_i &+& \omega_i^2 x_i =
  -\frac{\displaystyle \gamma}{\displaystyle m_i} \dot{x}_i,
   \qquad (i=1,2,\ldots,N) \label{nos1_1} \\
  \dot \gamma &=& \alpha \left(
  \frac{\displaystyle 2}{\displaystyle 3k_bTN}
  \sum_{i=1}^{N} \frac{\displaystyle m_i \dot{x}_i^2}{\displaystyle 2}
  -1 \right) \nonumber
\end{eqnarray}
It is important that the dissipative constant $\alpha$ is
suggested to be small, $\alpha \ll 1$, therefore, we may consider,
as the first approximation  the non dissipative regime for
which $\alpha \equiv 0$, $\gamma = 0$. In
this case there is an exact solution given by the equation
\begin{equation}
  x_i = A_i \cos{(\omega_i t + \phi_i)} ,  \quad
  \dot x_i = -A_i \omega_i \sin{(\omega_i t + \phi_i)} \label{xx}
\end{equation}
The energy of the $i$-th oscillator devided by mass $m$ reads
\begin{equation}
  e_i = \frac{\displaystyle \dot x_i^2}{\displaystyle 2} +
    \frac{\displaystyle \omega_i^2 x_i^2}{\displaystyle 2}
	  = \frac{\displaystyle A_i^2 \omega_i^2}{\displaystyle 2} \,
    \label{energy1}
\end{equation}
All the masses of the particles assumed equal, $m_i
= m$, oscillators differing by their frequencies $\omega_i$.
Equations (\ref{xx}), from the topological point of view, mean
that the ensemble's motion belongs to an $N$-dimensional torus,
the whole phase space being foliated by the tori. We shall
consider the system at temperature $T$. Since the parameter
$\alpha$ is small, we may take into account the nonlinear
dissipative terms on the right hand sides of Eqs. (\ref{nos1_1}),
within the framework of the averaging approach, that is by
substituting the basic equations (\ref{xx}) into the right hand
sides of the exact equations
\begin{eqnarray}
  \dot e_i &=& -\frac{\gamma}{m}
    \left(
        \frac{\displaystyle A_i^2 \omega_i^2}{\displaystyle 2} -
        \frac{\displaystyle A_i^2 \omega_i^2}{\displaystyle 2}
        \cos{(2\omega_i t + 2\phi_i)}
    \right)
  \nonumber \\
  \dot \gamma &=& \alpha
   \left(\frac{\displaystyle m}{\displaystyle 2T_0}
		\sum_{i=1}^{N} \frac{\displaystyle A_i^2 \omega_i^2}{\displaystyle 2}
		- 1
   \right)  \nonumber
\end{eqnarray}
and cancelling out the oscillating terms. Thus, we obtain the
averaged equations
\begin{eqnarray}
  \dot e_i &=& -\frac{\gamma}{m} \frac{\displaystyle A_i^2 \omega_i^2}{\displaystyle 2}
   \nonumber \\
  \dot \gamma &=& \alpha \left(\frac{\displaystyle m}{\displaystyle 2T_0}
  \sum_{i=1}^{N} \frac{\displaystyle A_i^2 \omega_i^2}{\displaystyle 2}
  - 1 \right)  \nonumber
\end{eqnarray}
or
$$
  \dot e_i = -\frac{\gamma}{m} e_i ,
  \qquad
  \dot \gamma = \alpha \left(\frac{\displaystyle m}{\displaystyle 2T_0}
  \sum_{i=1}^{N} e_i - 1 \right)
$$
as follows from Eq.(\ref{energy1}).

\par
\begin{center}
 \begin{figure}[h]
      \includegraphics*[width=3in, height=2in]{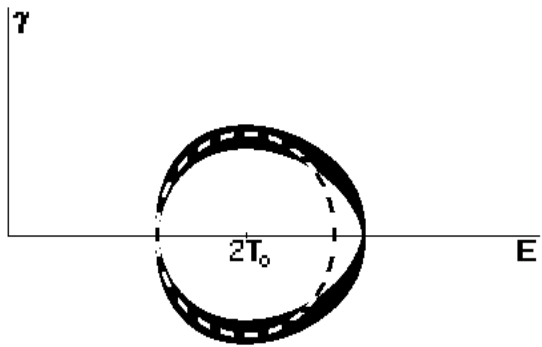}
     \caption{\label{oscil1p}
     Oscillations round the stationary solution, the characteristic
	 frequency $\sqrt{\alpha /m}$ . $N=1000   , \alpha = 0.01$,
	 $T_0 = Nk_bT/2$. Dashed line corresponds to
	 the averaged motion of $\gamma$ and $E$.
	 }
 \end{figure}
\end{center}

 Since
the total energy of the system is given by the equation
$$
  E = m \sum_{i=1}^{N} e_i ,
$$
we obtain the following two equations
$$
  \dot E = -\frac{\gamma}{m} E ,  \qquad
  \dot \gamma =
       \frac{\displaystyle \alpha}{\displaystyle 2T_0} (E-2T_0)
$$
which have the stationary solution:
$$
  E = 2T_0 , \qquad \gamma = 0,
$$
describing the oscillators at the temperature parameter
$$
   T_0 = \frac{\displaystyle k_b T N}{\displaystyle 2}
$$
Close to the stationary solutions we have the equations for energy
$$
      E = 2T_0 + Z
$$
in which $Z$ is small. Therefore the equations for $Z$ and
$\gamma$ acquire the form
$$
  \dot Z = -\frac{\gamma}{m} Z - 2\frac{\gamma}{m} T_0 , \qquad
  \dot \gamma = \frac{\displaystyle \alpha}{\displaystyle 2T_0}Z
$$
On linearizing the equations indicated above we obtain
$$
  \dot Z = - \frac{2 T_0}{m} \gamma, \qquad
  \dot \gamma = \frac{\displaystyle \alpha}{\displaystyle 2T_0}Z
$$
and hence the equation for $Z$
\begin{equation}
  \ddot Z + \frac{\alpha}{m} Z = 0   \label{Z}
\end{equation}
which has the form of  harmonic oscillator with the frequency
$$
  \sqrt{\frac{\alpha}{m}}
$$
Solutions to Eqs.(\ref{nos1_1}) are
illustrated in plane ($E$,$\gamma$), in FIG.(\ref{oscil1p})

We may use Eq.(\ref{Z}) for finding the time spent by the system
in the region of the phase-space corresponding to the energy
interval, $E_1 < E < E_2$; it is given by the equation

\begin{eqnarray}
  \Delta t &=& \frac{\displaystyle 2}{\displaystyle E_{max}-2T_0}
	   \left[  \arcsin \left(\frac{E_2-2T_0}{E_{max}-2T_0}
	   \sqrt{\frac{\displaystyle \alpha}{\displaystyle m}}\right)  - \right.
  \nonumber  \\
  &-&  \left. \arcsin  \left(\frac{E_1-2T_0}{E_{max}-2T_0}
	   \sqrt{\frac{\displaystyle \alpha}{\displaystyle m}}\right)
	   \right]    \label{arcsin}
\end{eqnarray}
which leads to the agreement with histogram of FIG.(\ref{hyst1p}).

\section{\label{sec:sum} Conclusion}

We have studied
the dynamics of an ensemble of harmonic oscillators confined to
the Nos\'e - Hoover thermostat, the number of the oscillators,
$N$, being large, and the dissipative constant, $\alpha$, small.
The numerical simulation, and the analytical averaging method,
indicate that the Nos\'e - Hoover model does not provide for {\em
thermodynamically meaningful distribution} of the system's samples
in the phase-space, e.g. there is no Gibbs' distribution. The
$\sqrt{\alpha / m}$-frequency law deserves some attention. Indeed,
if the mass $m$ of a particle, corresponding to the oscillator, be
$\approx 10^{-23} \; gr$, and $\alpha \sim 0.1 $, the frequency of
the oscillations generated by the thermostat dynamics, should be
in the region of $100$ GHz, that is low frequency region of
molecular vibrations. Therefore in applying the Nos\'e - Hoover thermostat
to simulating molecular dynamics in GHz - region, one might expect
spurious effects of parametric resonance.
At the same time it is worth noting that the Nos\'e - Hoover model
corresponds with a
hamiltonian system confined to dissipative environment, that is it
comprises a base hamiltonian system, e.g the oscillators, and a
dissipative extension formed by  ancila variables, in the present
case it is the variable $\gamma$. A similar systems, even though
more sophisticated, is the Leggett-Takagi theory of spin dynamics
in superfluid phases of helium-3, \cite{leggett}, in which the
equation describing the spin motion are augmented by an equation
for the order parameter, which contains a dissipative term. The
situation is reminiscent of that taking place in the hydrodynamics
treatment of viscous phenomena in the GHz - region, where according to the
theory worked out by Mandelstam and Leontovic, see \cite{LL}, the
effects of dissipation can be accommodated by employing an ancila
dynamical variable $\xi$, which describes certain states of the
system, e.g. concentration of a chemical reagent. The evolution
equations for $\xi$ have dissipative character, for they should
describe the system's coming to equilibrium, though the initial
equations for the system could be of hamiltonian form. The
Nos\'e - Hoover model may turn out to be of a similar kind and
thus  helpful in studying interesting physical problems.

\noindent {\bf Acknowledgment}

\noindent   We are thankful to A.Tenenbaum for the useful
correspondence.

\noindent This work was supported by the grants NS - 1988.2003.1,
and RFFI 01-01-00583, 03-02-16173, 04-04-49645.

  \vspace{2cm}

\end{document}